\documentclass[twocolumn,showpacs,amsmath,amssymb]{revtex4}
\usepackage[dvips]{graphics}
\def\imo{i}

\begin{document}
\title{Bifurcation of the quasinormal spectrum and Zero Damped Modes for rotating dilatonic black holes}
\author{K. D. Kokkotas}
\affiliation{Theoretical Astrophysics, Eberhard-Karls University of T\"{u}bingen, T\"{u}bingen 72076, Germany}
\author{R. A. Konoplya}\email{konoplya_roma@yahoo.com}
\affiliation{Institut f\"ur Theoretische Physik, Goethe-Universit\"at, Max-von-Laue-Str. 1, 60438 Frankfurt, Germany}
\author{A. Zhidenko}\email{olexandr.zhydenko@ufabc.edu.br}
\affiliation{Centro de Matem\'atica, Computa\c{c}\~ao e Cogni\c{c}\~ao, Universidade Federal do ABC (UFABC), Rua Aboli\c{c}\~ao, CEP: 09210-180, Santo Andr\'e, SP, Brazil}

\begin{abstract}
It has been recently found that for the near extremal Kerr black holes  appearing of Zero Damped Modes (accompanied by qusinormal mode branching) signifies about inapplicability of the regime of small perturbations and the onset of turbulence. Here we show that this phenomena is not limited by Kerr or Kerr-Newman solutions only, but also takes place for rotating dilatonic black holes for which we have found Zero Damped Modes both numerically and analytically. We have also shown that, contrary to recent claims, there is no instability of a charged massive scalar field in the background of the rotating dilatonic black hole under physically adequate boundary conditions. Analytic expression for dominant quasinormal frequencies is deduced in the regime of large coupling $q Q$, where $q$ and $Q$ are the field and black hole charges respectively.
\end{abstract}
\maketitle

\section{Introduction}

A remarkable feature of the characteristic (quasinormal) oscillation spectra of black holes has been recently observed in \cite{Yang:2012pj}:
in the regime of high black hole rotation a kind of the modes branching occurs. One of the branches consist of the modes which for the Kerr black hole are given by the following formula
\begin{equation}
\omega M \approx \frac{m}{2} - A(\ell, m) \sqrt{\epsilon} - \imo \left(n+\frac{1}{2}\right) \sqrt{\frac{\epsilon}{2}},
\end{equation}
where $\epsilon=\sqrt{1-a/M}$ vanishes in the limit of the extreme rotation. Thus, when approaching extreme rotation, quasinormal modes with vanishing damping rate, that is, very long and asymptotically infinite lifetime appear. These modes were called Zero Damped Modes (ZDMs) and the real oscillation frequencies of these modes were found in \cite{Hod:2008zz} under assumption of smallness of their damping rates.

Existence of such long living modes which were obtained initially within formalism of small perturbations of black holes might mean that regime of small perturbation does not describe near extremal black hole adequately anymore. Indeed, as it has been recently shown in \cite{Yang:2014tla}, ZDMs signify a qualitatively new, turbulent state of the Kerr black hole. Moreover, if the black hole is described by the Kerr solution and may achieve near extremal rotations, observational consequences in the gravitational wave structure and accreting processes become at least theoretically plausible \cite{Yang:2014tla}.

The modes branching as well as accurate calculation of ZDMs were extended to Einstein-Maxwell system (Kerr-Newman solution) \cite{Konoplya:2013rxa,Mark:2014aja}. An appealing question is whether the phenomena of Zero Damped Modes (and thereby of the black hole turbulence) occurs only for Kerr black holes may be answered after understanding the reason inducing such a peculiar behavior of the spectrum. Therefore, here we shall try a straightforward way and will see if the phenomena of ZDMs in the near extremal regime is not limited by the Kerr (or Kerr-Newman) solution, and takes place also for Sen black holes \cite{Sen:1992ua}. We shall show that the both the bifurcation and ZDMs do exist for Sen black hole and, unlike the Kerr-Newman case \cite{Konoplya:2013rxa,Mark:2014aja}, the phenomena take place also for not small electric charge $Q$.

Another motivation for analysis of quasinormal spectra of the charged scalar field in the rotating dilatonic background is a recent statement that the charged field is unstable in its background \cite{Siahaan:2015xna}. Here we shall show that this claim is not related to fields's (in)stability in its usual sense, but is related to the bound states of the field. Here we have also calculated fundamental quasinormal frequencies and found analytical formula for $\omega$ in the regime of large $q Q$.

The paper is organized as follows in Sec II we give the basic formulas for the dilatonic metric for the Sen black hole \cite{Sen:1992ua}
including the separation of variables for the scalar field equation in the black hole background. Sec. III is devoted to numerical analysis of qusinormal modes with the help of the Frobenius method and deduction of the analytic expression for QNMs in the regime of large $q Q$. Sec. IV discusses effects of spectrum bifurcation and Zero Damped Modes, including deduction of analytical formula for ZDMs. Sec. V comments on superradiance and stability of a massive charged scalar field.

\section{Scalar field in the rotating dilatonic black hole background}

The Sen black hole \cite{Sen:1992ua} (axion-dilaton black hole with zero NUT \cite{Garcia:1995}) is given by the following line element \cite{Okai:1994td}
\begin{eqnarray}
ds^2 &=& \frac{\Delta_r}{\Sigma}(dt-a\sin^2\theta d\varphi)^2-\Sigma
\left(\frac{dr^2}{\Delta_r}+d\theta^2\right)\\\nonumber
&&-\frac{\Delta_\theta \sin^2\theta}{\Sigma}
[adt-(r^2+2br+a^2)d\varphi]^2,
\end{eqnarray}
where
\begin{eqnarray}\nonumber
\Delta_r&=&r^2-2(M-b)r+a^2,\\
\Sigma&=&r^2+2br+a^2\cos^2\theta,\nonumber
\end{eqnarray}
$a$ is the rotation parameter, $M$ is the ADM mass and $M>b\geq0$. The Maxwell and dilaton fields are given by
\begin{eqnarray}
A_{\mu}dx^{\mu}&=&Q\frac{r}{\Sigma}(dt-a\sin^2\theta d\varphi),\\
e^{2\phi}&=&W\frac{r^2+a^2\cos^2\theta}{\Sigma},
\end{eqnarray}
where the electric charge $Q$ is related to $b$ as
$$Q^2=2WMb.$$

We shall parameterize the metric by the following three nonnegative parameters: the event horizon $r_+$, the inner horizon $r_-$, and the parameter $b$,
$$0\leq r_- < r_+.$$
Then, the black hole's mass and rotation parameter can be written as
$$2M=r_++r_-+2b,\qquad a^2=r_+r_-.$$

A massive charged scalar field satisfies the Klein-Gordon equation,
\begin{equation}\label{KG}
\left(\frac{1}{\sqrt{-g}}D_{\alpha}g^{\alpha\beta}\sqrt{-g}D_{\beta}+\mu^2\right)\psi=0,
\end{equation}
where
$$D_{\alpha}\equiv\frac{\partial}{\partial x^\alpha}-iq A_{\alpha},$$
$q$ and $\mu$ are the field's charge and mass respectively.

One can separate variables by the ansatz \cite{Wu:2001xh}
\begin{equation}\label{anzats}
\psi = e^{-\imo \omega t + \imo m \phi} S(\theta) R(r),
\end{equation}
where $S(\theta)$ obeys the following equation
\begin{eqnarray}\label{angularpart}
\left(\frac{d^2}{d \theta^2} + \cot \theta \frac{d}{d\theta} - \frac{m^2}{\sin^2 \theta} - a^2 \omega^2\sin^2 \theta\right.&&
\\\nonumber\left. + 2 m a \omega + \lambda - \mu^2 a^2 \cos^2 \theta \right) S(\theta) &=& 0,
\end{eqnarray}
and $\lambda$ is the separation constant.

This equation can be solved numerically for any value of $\omega$ in the same way as the equation for a massive scalar field in the Kerr black-hole background \cite{Konoplya:2006br}. Let us note that, when $\mu=0$, Eq.~(\ref{angularpart}) can be reduced to the well-known equation for the spheroidal functions. In this case, for any fixed value of $\omega$ the separation constant $\lambda$ can be found numerically using the continued fraction method \cite{Suzuki:1998vy}. When the effective mass is not zero, the separation constant $\lambda(\omega, \mu)$ can be expressed, in terms of the eigenvalue for spheroidal functions $\lambda(\omega)$ \cite{Kokkotas:2010zd}, as
$$\lambda(\omega,\mu)=\lambda(\sqrt{\omega^2-\mu^2},0)+2ma(\sqrt{\omega^2-\mu^2}-\omega)+\mu^2a^2.$$
When $a=0$, one has $\lambda=\ell(\ell+1),~\ell=0,1,2\ldots$. For nonzero values of $a$, the separation constant can be enumerated by the integer multipole number $\ell\geq|m|$.

\section{Quasinormal frequencies and the Frobenius method}

The radial function $R(r)$ satisfies the following wave equation \cite{Siahaan:2015xna}
\begin{equation}\label{radialpart}
\left(\frac{d}{d r}\Delta_r\frac{d}{d r}+\frac{K^2}{\Delta_r}-\mu^2(r^2+2br)-\lambda\right)R(r)=0,
\end{equation}
where
$$K=\omega(r^2+2br+a^2)-ma-qQ r.$$

Equation (\ref{radialpart}) has an irregular singularity at spatial infinity and two  regular singularities at $r=r_\pm$.
We impose the quasinormal boundary conditions, implying purely outgoing wave at spatial infinity and purely ingoing wave at the event horizon $r_+$.
Then, the appropriate Frobenius series is
\begin{equation}
R(r)=\left(\frac{r-r_+}{r-r_-}\right)^{-2\imo\alpha}e^{\imo \rho}(r-r_-)^{\imo\sigma}y(r),
\end{equation}
where
$$\alpha=\frac{K}{d\Delta_r/dr}\Biggr|_{r=r_+},\qquad \frac{d\rho}{dr}=\frac{K}{\Delta_r}+\sqrt{\omega^2-\mu^2}-\omega,$$
and $\sigma$ can be find from the regularity condition of the equation for $y$ at spatial infinity.

For the QNMs the function $y(r)$ is regular at the horizon and spatial infinity, so that the series in the vicinity of the event horizon
$$y(r)=\sum_{k=0}^{\infty}a_k(\omega)\left(\frac{r-r_+}{r-r_-}\right)^k$$
converge at $r=\infty$ \cite{Leaver:1985ax}.

The coefficients $a_k(\omega)$ satisfy the three-terms recurrence relation, which has the form
\begin{eqnarray}
&&\alpha_0a_1+\beta_0a_0=0\,, \nonumber \\
&&\alpha_ka_{k+1}+\beta_ka_k+\gamma_ka_{k-1}=0\,,
\ k=1,2,\ldots.
\end{eqnarray}
The coefficients $\alpha_k$, $\beta_k$, and $\gamma_k$ can be found in a closed form. For the massless case ($\mu=0$) the corresponding expressions take the following form
\begin{eqnarray}\nonumber
\alpha_k&=&(k+1)\Biggr(2\imo(\omega r_+(r_++r_-+2b)-qQr_+-am)\\&&-(k+1)(r_+-r_-)\Biggr),\label{ak}\\
\beta_k&=&4 a^2 qQ \omega-8r_+(b+r_+) (r_+ + r_- + 2b) \omega^2\nonumber\\
   &&+\imo (3r_+-r_-) (1+2k) (qQ-2b\omega)\nonumber\\
   &&+8 (am +qQr_+)(b +r_+)\omega\label{bk}\\
   &&+2 \imo(1+2k)(a m-2r_+^2 \omega)\nonumber\\
   &&+(2 k^2+2k+1+\lambda) (r_+-r_-) \nonumber\\
   &&+4 qQ r_+(r_++2b)\omega-4qQ(am+qQr_+),\nonumber\\
\gamma_k&=&-\Biggr(k-2\imo(\omega(r_++r_-+2b)-qQ)\Biggr)\Biggr(k(r_+-r_-)\nonumber\\
   &&-2\imo(\omega r_+(r_++r_-+2b)-qQr_+-am)\Biggr).\label{ck}
\end{eqnarray}

\begin{figure*}
\resizebox{\linewidth}{!}{\includegraphics*{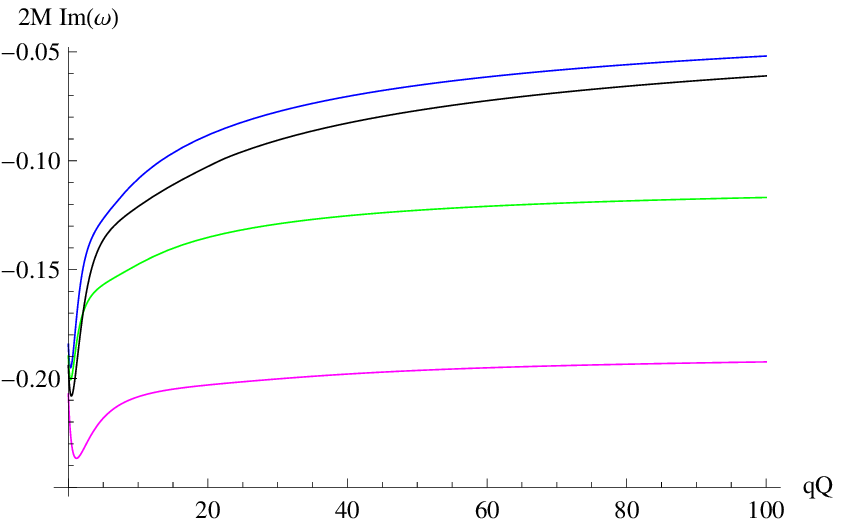}\includegraphics*{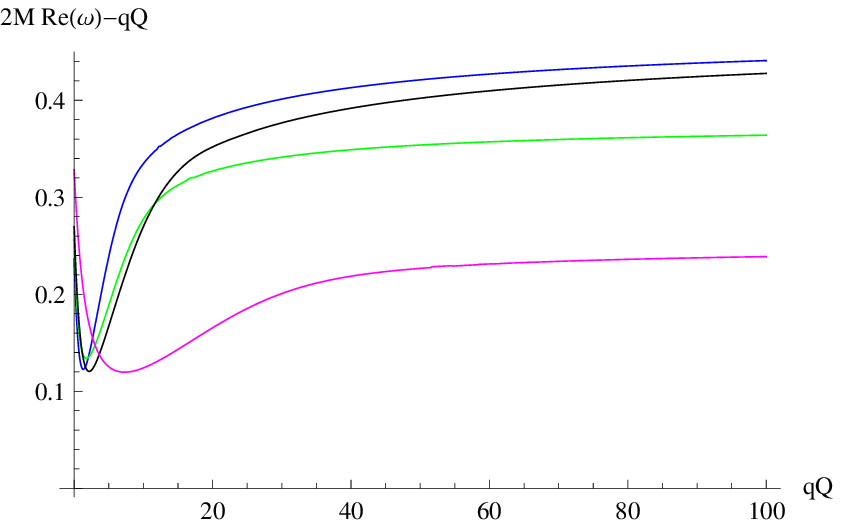}}
\caption{Fundamental quasinormal modes $\ell =m=0$ for various values of $b$ and $a$ (from top to bottom): $b=0.25r_+, ~a=0.95r_+$ (blue), $b=r_+,~a=0.95r_+$ (black), $b=0.5r_+, ~a=0.75r_+$ (green), $b=2r_+, ~a=0.5r_+$ (magenta). In the units of mass the asymptotical behaviour does not depend on $b$: blue and black lines approach the same limit.}\label{fundqnmqQ}
\end{figure*}

\begin{figure*}
\resizebox{\linewidth}{!}{\includegraphics*{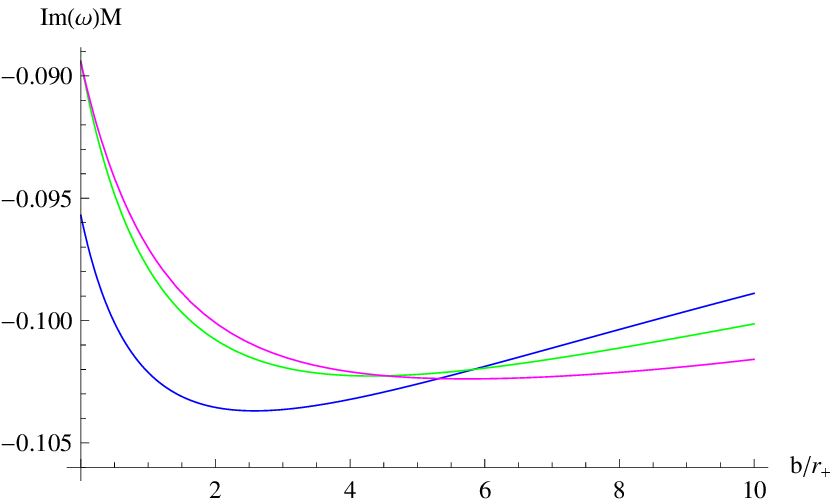}\includegraphics*{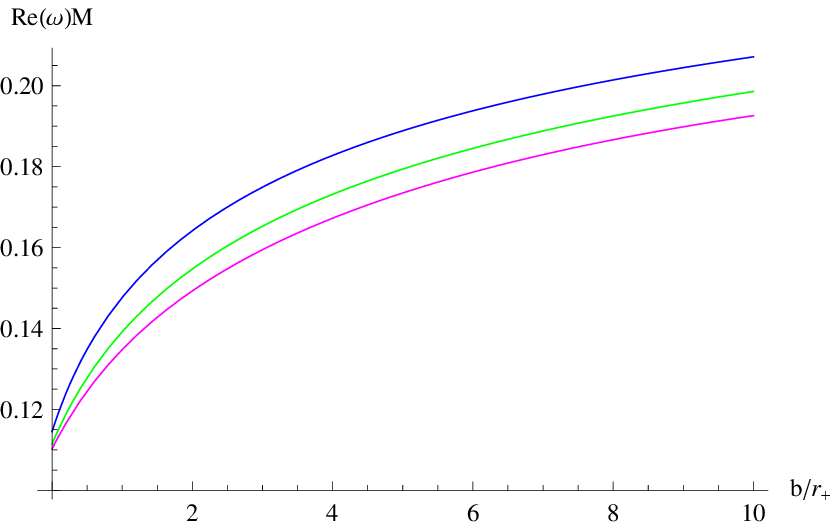}}
\caption{Fundamental quasinormal modes $\ell=m=0$ for various values of $a$ (for large $b$ from top to bottom): $a=0.5r_+$ (blue), $a=0.75r_+$ (green), $a=0.95r_+$ (magenta).}\label{fundqnm}
\end{figure*}

In order to find QNMs we solve numerically the following equation \cite{Nollert,Zhidenko:2006rs}
\begin{eqnarray}
0=\beta_0-{\alpha_0\gamma_1\over\beta_1-}
{\alpha_1\gamma_2\over\beta_2-}{\alpha_2\gamma_3\over\beta_3-}
\cdots \,.
\label{a-eq}
\end{eqnarray}
For the contour plots we calculate logarithm of the absolute value of the righthand side of the above equation (\ref{a-eq}). For more details on Frobenius method see Sec.~3 of \cite{Konoplya:2011qq}.

In order to find the overtone number $n$ we rewrite (\ref{a-eq}) as
\begin{eqnarray}\label{continued_fraction} \beta_n-\frac{\alpha_{n-1}\gamma_{n}}{\beta_{n-1}
-}\frac{\alpha_{n-2}\gamma_{n-1}}{\beta_{n-2}-}\frac{\alpha_{n-3}\gamma_{n-2}}{\beta_{n-3}-}\ldots\frac{\alpha_0\gamma_1}{\beta_0}=\qquad\\\nonumber
\frac{\alpha_n\gamma_{n+1}}{\beta_{n+1}-}\frac{\alpha_{n+1}\gamma_{n+2}}{\beta_{n+2}-}\frac{\alpha_{n+2}\gamma_{n+3}}{\beta_{n+3}-}\ldots,
\end{eqnarray}
and search for the most stable root.

Following \cite{Konoplya:2013rxa}, we observe that for, large $qQ$, $\alpha_n={\cal O}(qQ)$, $\beta_n={\cal O}(qQ)^2$, and $\gamma_n={\cal O}(qQ)^2$, implying that~(\ref{continued_fraction}) reads
\begin{equation}\label{continued_fraction_qQ}
\frac{\beta_n}{(qQ)^2} = {\cal O}\left(\frac{1}{qQ}\right).
\end{equation}

Considering $\lambda\ll (qQ)^2$, from (\ref{continued_fraction_qQ}) we find that
\begin{equation}
\forall n:\qquad\lim_{qQ\to\infty}\frac{\omega}{qQ}=\frac{1}{r_++r_-+2b}.
\end{equation}
In this regime, $\beta_n\gg\alpha_n={\cal O}(1)$, $\beta_n\gg\gamma_n={\cal O}(1)$, and, as shown in \cite{Konoplya:2013rxa},
\begin{equation}
\lambda=2a\omega\left(1+2\left[\frac{\ell-m}{2}\right]\right)+{\cal O}(1),
\end{equation}
where the brackets denote the integer part. Then (\ref{continued_fraction}) is reduced to
\begin{equation}
\frac{\beta_n}{qQ}={\cal O}\left(\frac{1}{qQ}\right),
\end{equation}
from which we find the subdominant contributions to the QNMs for $qQ\gg1$,
\begin{eqnarray}\nonumber
2M\omega&=&\omega(r_++r_-+2b)=qQ+\frac{a}{r_+}\left(\left[\frac{\ell+m}{2}\right]+\frac{1}{2}\right)\\
&&-\imo\left(1-\frac{a^2}{r_+^2}\right)\frac{2n+1}{4}+{\cal O}\left(\frac{1}{qQ}\right).\label{asymptotw}
\end{eqnarray}
The above formula is in excellent agreement with numerical data on fig.~\ref{fundqnmqQ}.

We have also calculated the fundamental quasinormal modes ($\ell = m=0$) for various parameters of rotation as a function of $b$ (see fig.~\ref{fundqnm}). There it can be seen that increasing of the dilaton parameter $b$ from $0$ up to value of a few $r_{+}$ leads to considerable decreasing of the damping rate by quite a few times, so that dilatonic black hole has much longer living modes for moderate values of $b$. Then, for larger $b$ the damping rate slowly increases, apparently achieving some constant asymptotically.  The real oscillation frequencies monotonically decrease and also seem to approach constant values in the regime of large $b$.

\section{Bifurcation and Zero Damped Mode: Numerical and Analytical Treatments}

\begin{figure*}
\resizebox{\linewidth}{!}{\includegraphics*{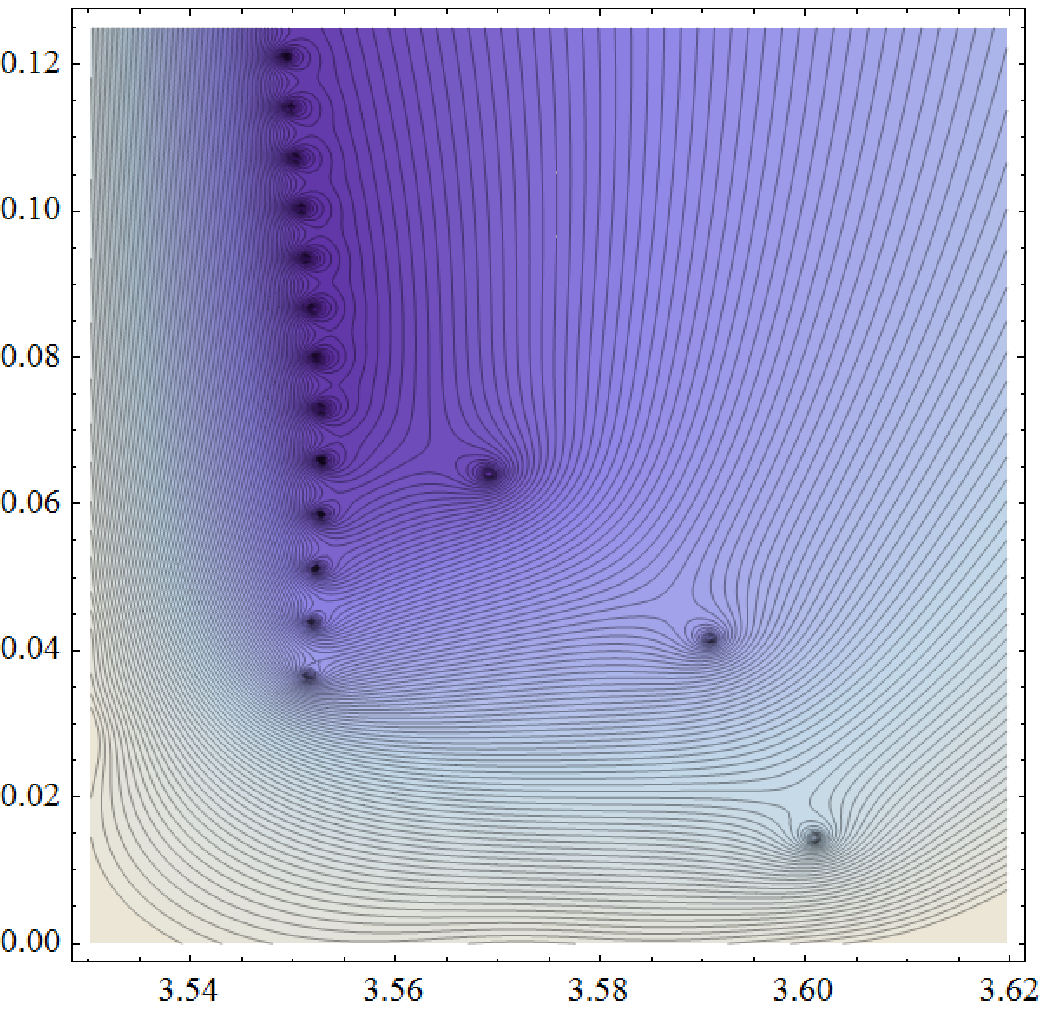}\includegraphics*{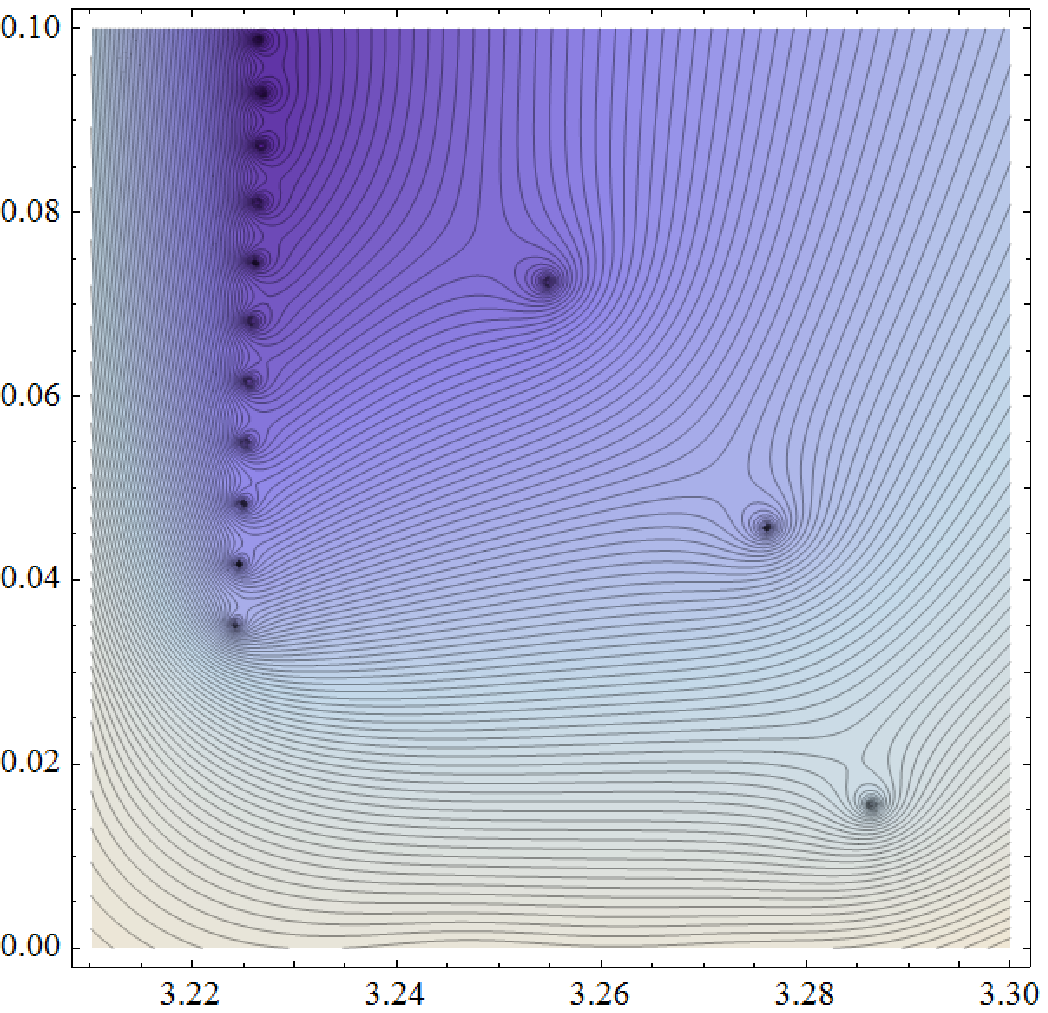}\includegraphics*{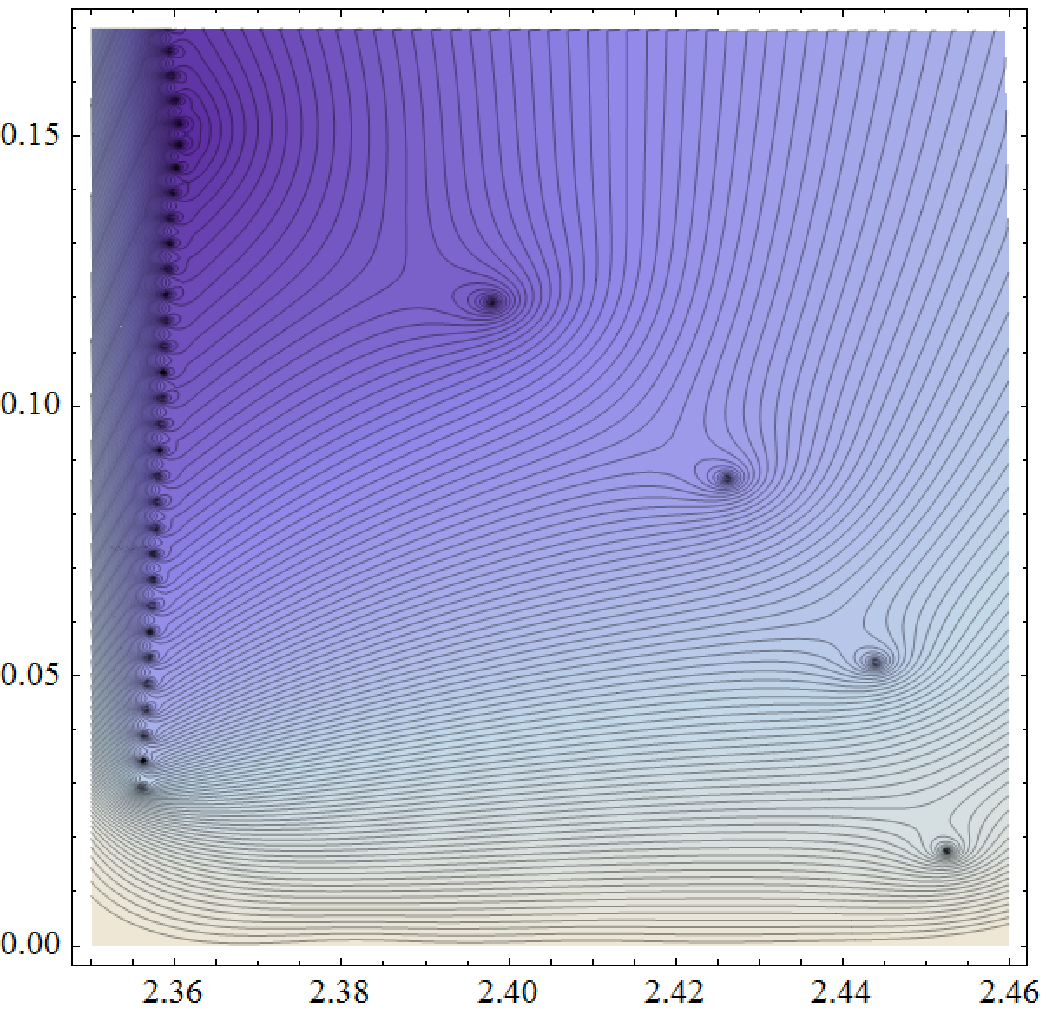}}
\caption{Level plots of the logarithm of the absolute value of the difference between the left and right sides of the equation
with continued fraction as a function of the real (horizontal axis) and imaginary (vertical axis) parts of the frequency
for the scalar field $\ell=10$, $m=7$, $a=0.985957r_+$ ($r_-=0.972111r_+$), from left to right: $b=0$, $b=0.1r_+$, $b=0.5r_+$.}\label{bplot}
\end{figure*}

\begin{figure*}
\resizebox{\linewidth}{!}{\includegraphics*{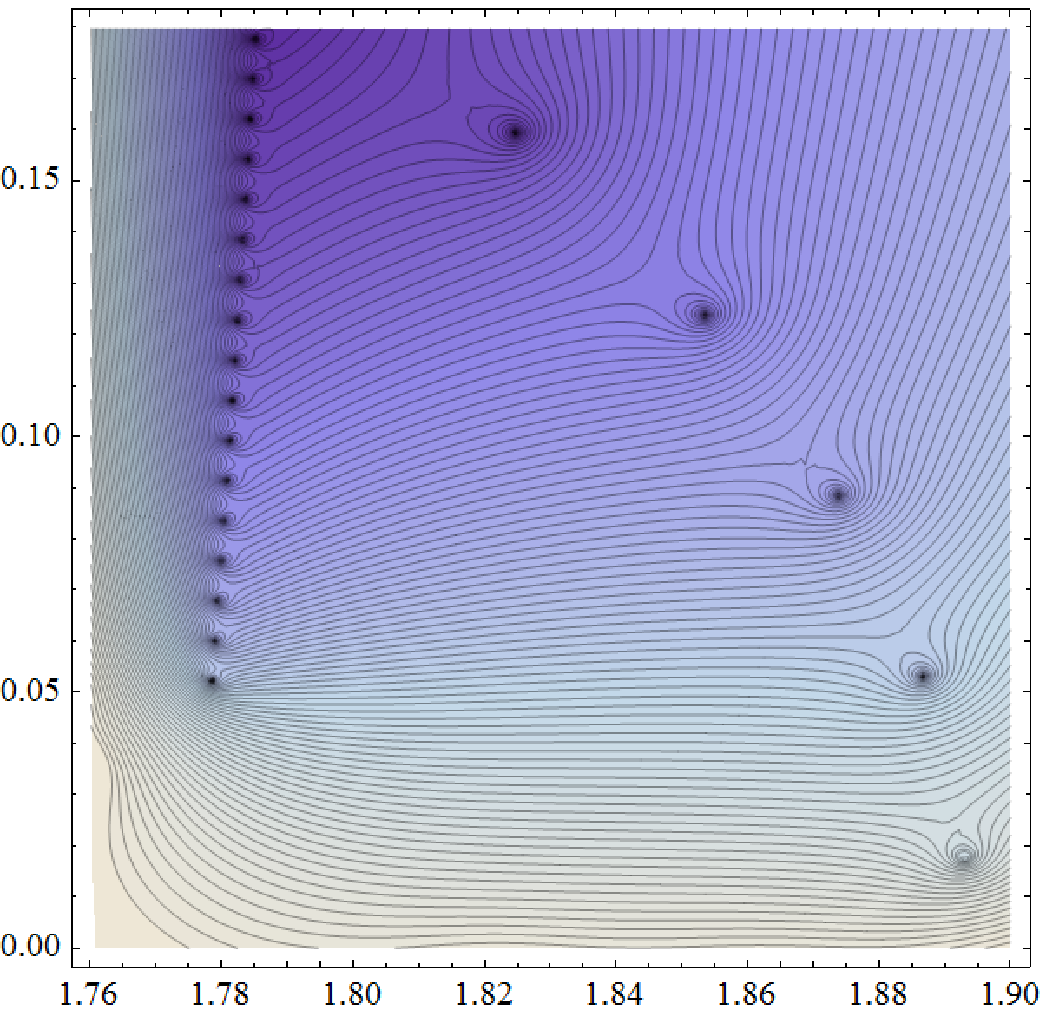}\includegraphics*{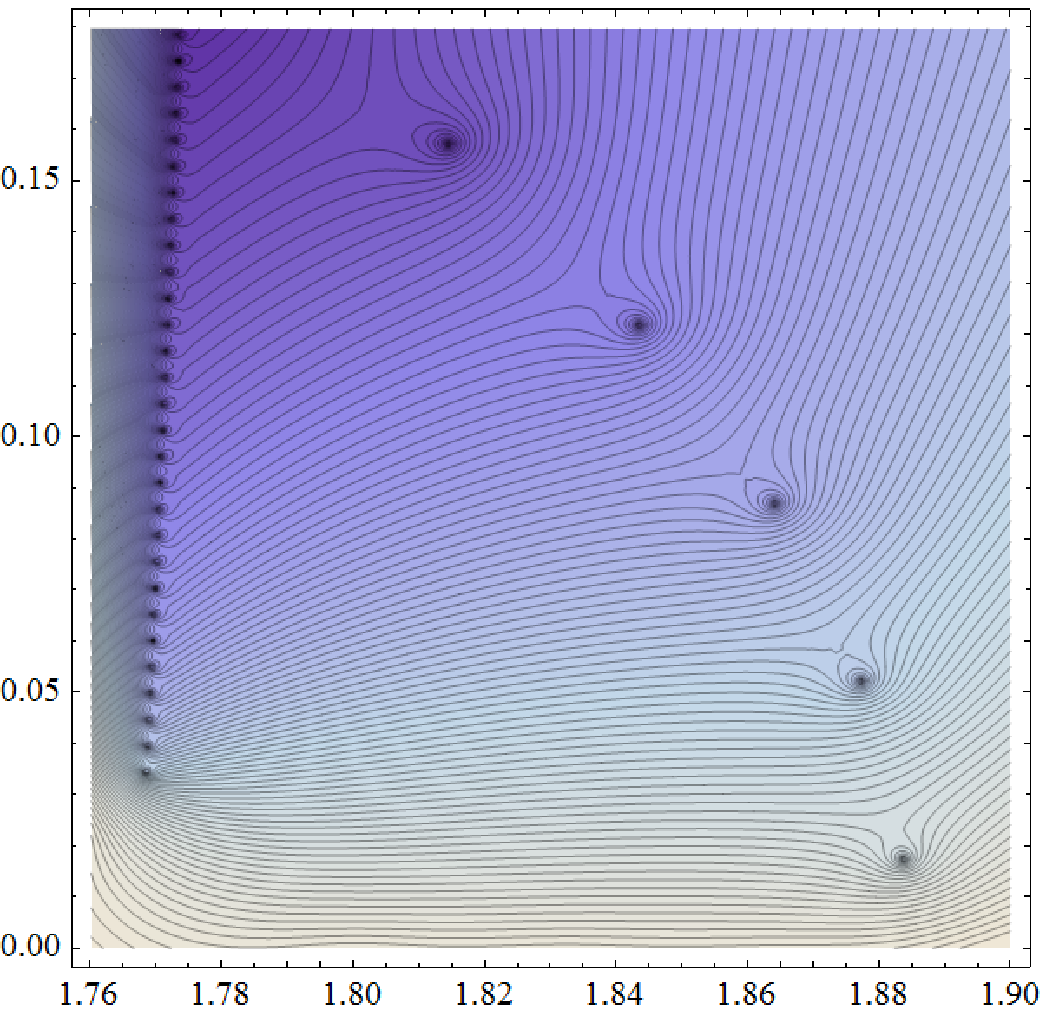}\includegraphics*{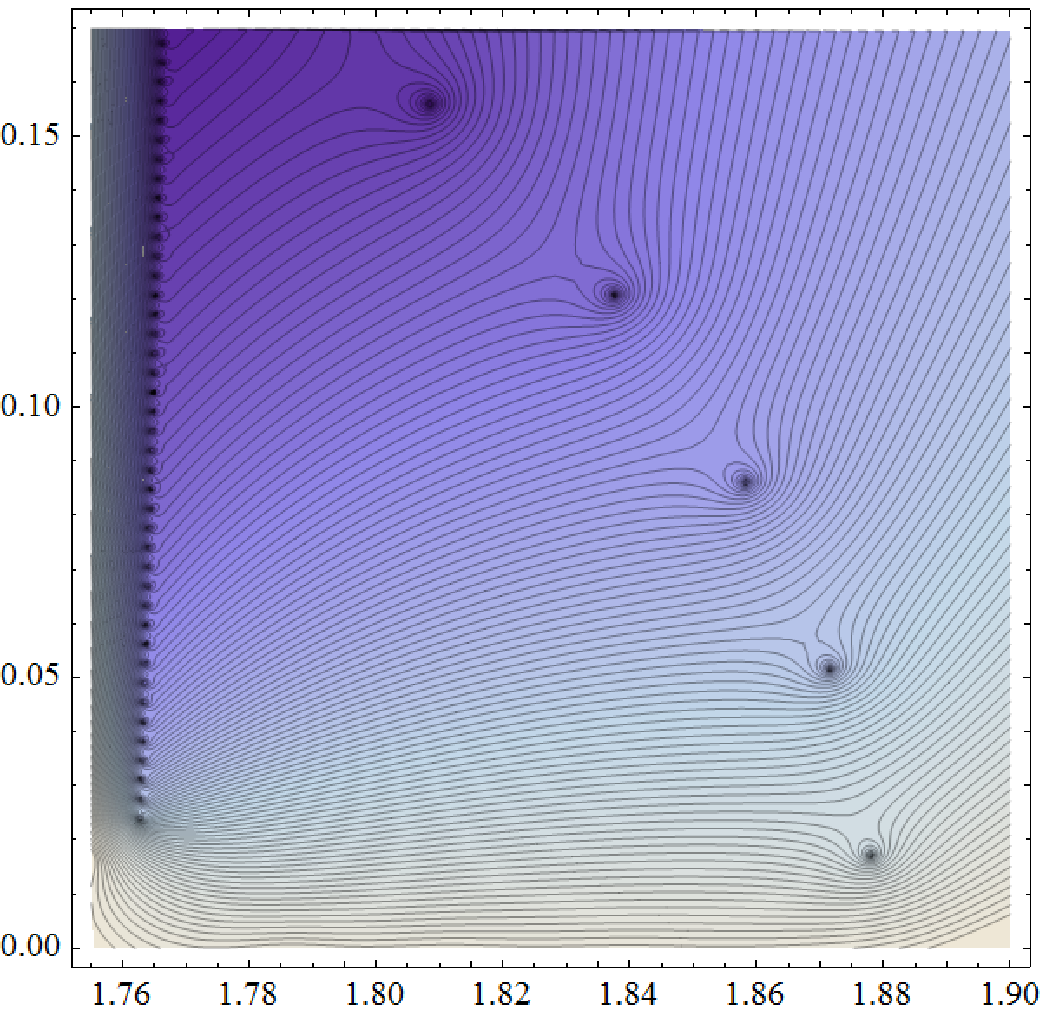}}
\caption{Level plots of the logarithm of the absolute value of the difference between the left and right sides of the equation
with continued fraction as a function of the real (horizontal axis) and imaginary (vertical axis) parts of the frequency
for the scalar field $\ell=10$, $m=7$, $b=r_+$, from left to right: $a=0.97r_+$, $a=0.98r_+$, $a=0.985957r_+$.}\label{b=1}
\end{figure*}

The smaller the multipole number $\ell$ the higher is "critical" rotation at which the bifurcation takes place. Therefore, in order to demonstrate the effect, it is reasonable to choose some high values of $\ell$ and non-zero $m$. On figs.~\ref{bplot},~\ref{b=1} it is shown the logarithm of the absolute value of difference between the left and right hand sides of the continued fraction equation (\ref{a-eq}) as a function of real and imaginary parts of $\omega$. On figs.~\ref{bplot},~\ref{b=1} the left branches correspond to the so-called Zero-Damped modes (ZDMs) which have been found for Kerr black holes in \cite{Yang:2012pj}. ZDM-branch asymptotically approaches pure real modes in the limit of extreme rotation.

For Kerr-Newman black holes \cite{Konoplya:2013rxa}, for small values of black hole's charge $Q$ the mode branching occurs owing to extremal rotation and not owing to extremal charge. Larger values of charge $Q$ break down the bifurcation effect. Remarkably, for the Sen black hole, the mode branching occurs even for not small values of charge in the near extremal limit.

In \cite{Yang:2012pj} it was shown that ZDMs satisfy $\alpha\to0$ as $a\to r_+$. Acting in similar fashion with \cite{Yang:2012pj}, we conclude that ZDMs have the following form
\begin{equation}
\omega = \frac{m+qQ}{2(r_++b)}+{\cal O}\left(\sqrt{1-\frac{a}{r_+}}\right).
\end{equation}
Quasinormal modes depicted on our plots (figs. \ref{bplot}, \ref{b=1}) is in complete concordance with the above expression for $\omega$.
Non-ZDMs do not scale as $\propto(r_++b)^{-1}$ leading to larger separation between ZDMs and non-ZDMs for higher values of $b$ (see Fig.~\ref{bplot}).

\section{Remark on stability and superradiance}

Recently, it was claimed in \cite{Siahaan:2015xna} that a massive charged scalar field in the background of Sen black hole has unstable modes, which might be also important for the black hole stability. As previous publications on quasinormal modes of charged scalar fields in various asymptotically flat or de Sitter black hole backgrounds, including Kerr-Newman and dilaton ones showed no unstable modes \cite{Kokkotas:2010zd,raznoe,Konoplya:2013rxa},   this motivated us to consider here also the dominant quasinormal modes in the range of parameters, according to \cite{Siahaan:2015xna}, where the instability could be expected, namely, the range determined by the superradiant regime.

To show the range of frequencies that gives rise to the superradiance effect, we need to study the asymptotic behavior of radial solution $R(r)$ at $r\to r_+$ and $r\to \infty$. Taking $r\to \infty$ the general solution to (\ref{radialpart}) is a superposition of ingoing and outgoing waves
\begin{eqnarray}\label{Y-far}
R &=& \exp\left(-\imo r\sqrt {\omega ^2  - \mu ^2 }+{\cal O}(ln(r))\right)\\\nonumber&&+{\cal R}\exp\left(\imo r\sqrt{\omega ^2  - \mu ^2}+{\cal O}(ln(r))\right),
\end{eqnarray}
where ${\cal R}$ is the reflection coefficient.

There is only ingoing wave near the horizon
\begin{equation}\label{Y-near}
R = \left(r-r_+\right)^{-\imo\alpha}\left({\cal T}+{\cal O}(r-r_+)\right),
\end{equation}
where ${\cal T}$ is the transmission coefficient.

It is possible to prove for real frequencies that ${\cal R}>1$ only if $\displaystyle\frac{\alpha}{\omega}<0$, in what follows that \cite{Siahaan:2015xna}
\begin{equation}\label{superrads-freq}
0 < \omega  < \frac{ma+qQ r_+}{2Mr_+}\quad\mbox{or}\quad\frac{ma+qQ r_+}{2Mr_+}<\omega<0.
\end{equation}
It is worthwhile of notice that the superradiance condition for the Sen black hole coincides with the Kerr black hole, but differs for the Kerr-Newman black hole (cf.,~\cite{Konoplya:2013rxa}).

\begin{figure}
\resizebox{\linewidth}{!}{\includegraphics*{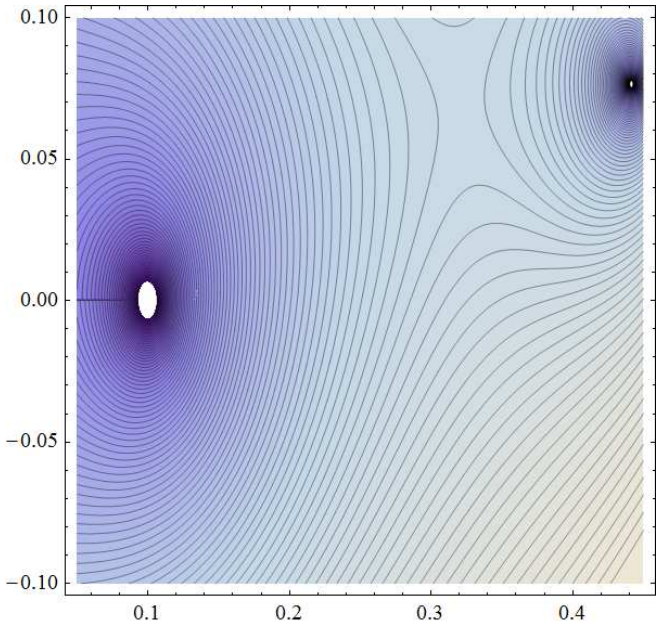}}
\caption{$\ell=m=1$, $\mu r_+=0.1$, $a=b=0.5r_+$, $q Q=0.1$, $\omega r_+ = 0.44148 - 0.07654\imo$ is the fundamental mode, $\omega=\mu$ is a local minimum which can be approached as close as we want. The requirement $\omega \leq \mu$ along the real axis does not allow to impose the quasinormal boundary conditions.}\label{massive-field}
\end{figure}

We have made thorough numerical search of quasinormal frequencies in the above range of parameters and found no unstable quasinormal modes (an example of non-existence of the unstable mode, expected in~\cite{Siahaan:2015xna} can be seen in fig.~\ref{massive-field}).
Therefore, no any true instability occurs for the quasinormal spectrum of a massive charged scalar field in the Sen background, contrary to \cite{Siahaan:2015xna}. The reason of this discrepancy is straightforward: boundary conditions used in \cite{Siahaan:2015xna} is not the one, corresponding to black hole's response to the initial perturbation, that is, not the quasinormal mode boundary condition. The latter implies purely ingoing waves at the event horizon and purely outgoing waves at infinity, while the boundary conditions in \cite{Siahaan:2015xna} correspond to bound states of the field \cite{Li:2015mqa}. Instability of such bound states is known for a long time \cite{Detweiler:1980uk}, occurs for a massive field in the rotating black hole's background and simply means that the corresponding particle will change its quantum state in order to move from the superradiant sector. Thus, bound states of a neutral massive field are unstable around Kerr black hole in the regime of superradiance \cite{Detweiler:1980uk}, but no instability takes place under quasinormal boundary condition \cite{Konoplya:2006br}. Generally speaking, superradiance do can lead to the instability in the classical sense, but these are cases of asymptotically non-flat backgrounds, such as, for example, asymptotically de Sitter background \cite{Konoplya:2014lha}. Notice, that in the latter case, no rotation is necessary to trigger the instability.

\section{Conclusion}

We have shown that the effect of bifurcation of the quasinormal spectra and appearance of Zero Damped Modes takes place not only for Kerr, but also for Sen black holes. Unlike for the Kerr-Newman black hole, even highly charged black holes have Zero Damped Modes in the limit of extreme rotation.

This means that inapplicability of the regime of linear perturbations for the near extreme black hole's rotation is not a peculiar property of the Kerr solution, but occurs also for other models of rotating axisymmetric black holes. Apparently, the effects of ZDMs and the bifurcation occurs for much larger class of geometries than Kerr-Newman or Sen black holes, and further research may find possible connections between these phenomena and black hole geometries.

We have also shown that, contrary to the recent claim \cite{Siahaan:2015xna}, quasinormal spectrum of a charged scalar field in the background of the rotating dilatonic black hole does not have unstable modes. Analytical expressions for Zero Damped Modes (in the near extremal regime) and for ``ordinary" quasinormal modes (in the regime of high $q Q$) were deduced. Numerical data for QNMs are in excellent agreement with the analytic formulas.

\section*{Acknowledgments}
A.~Z. was supported by Conselho Nacional de Desenvolvimento Cient\'ifico e Tecnol\'ogico (CNPq).
At the initial stage of this work R. A. K. was supported by the Alexander von Humboldt Alumni Program, and at the final part by
the BlackHoleCam Synergy Project. R. A. K. acknowledges also hospitality of Theoretical Astrophysics Group of  Eberhard-Karls University in T\"{u}bingen.

\end{document}